\documentclass[%
 reprint,
superscriptaddress,
 amsmath,amssymb,
 aps,
]{revtex4-1}
\usepackage[normalem]{ulem}

\usepackage{graphicx}
\usepackage{color}
\usepackage{dcolumn}
\usepackage{bm}

\begin{document}

\preprint{APS/123-QED}

\title{Phonon order and reststrahlen  bands of polar vibrations in crystals with monoclinic symmetry}

\author{Mathias Schubert}
\email{schubert@engr.unl.edu}
\homepage{http://ellipsometry.unl.edu}
\affiliation{Department of Electrical and Computer Engineering, University of Nebraska-Lincoln, Lincoln, Nebraska 68588, U.S.A.}
\affiliation{Leibniz Institute for Polymer Research, Dresden, Germany}
\affiliation{Terahertz Materials Analysis Center, Department of Physics, Chemistry and Biology, Link{\"o}ping University, SE 58183, Link{\"o}ping, Sweden}
\affiliation{Center for III-Nitride Technology C3NiT - Janz{\'e}n, Link{\"o}ping University, SE 58183, Link{\"o}ping, Sweden}
\author{Alyssa Mock}
\affiliation{Terahertz Materials Analysis Center, Department of Physics, Chemistry and Biology, Link{\"o}ping University, SE 58183, Link{\"o}ping, Sweden}
\affiliation{Center for III-Nitride Technology C3NiT - Janz{\'e}n, Link{\"o}ping University, SE 58183, Link{\"o}ping, Sweden}
\author{Rafa{\l} Korlacki}
\affiliation{Department of Electrical and Computer Engineering, University of Nebraska-Lincoln, Lincoln, Nebraska 68588, U.S.A.}
\author{Vanya Darakchieva}
\affiliation{Terahertz Materials Analysis Center, Department of Physics, Chemistry and Biology, Link{\"o}ping University, SE 58183, Link{\"o}ping, Sweden}
\affiliation{Center for III-Nitride Technology C3NiT - Janz{\'e}n, Link{\"o}ping University, SE 58183, Link{\"o}ping, Sweden}

\date{\today}

\begin{abstract}
We derive from an eigendielectric displacement vector summation approach the frequency order of the polar phonon modes and the polarization-dependent structure of the reststrahlen bands within the monoclinic plane of materials with monoclinic crystal symmetry. We show that phonon modes occur in associated pairs of transverse and longitudinal optical modes, and that pairs either belong to inner or outer phonon modes. Inner mode pairs are nested within outer mode pairs. Outer mode pairs cause polarization-dependent reststrahlen bands. Inner mode pairs cause polarization-independent reststrahlen bands. The directional limiting frequencies within the Born-Huang approach are bound to within outer mode regions not occupied by inner mode pairs. Hence, an unusual phonon mode order can occur where both lower-frequency as well as upper-frequency limits for the directional modes can be both transverse and/or longitudinal modes. We exemplify our findings using experimental data for the recently unraveled case of monoclinic symmetry $\beta$-Ga$_2$O$_3$ [Phys. Rev. B~\textbf{93}, 125209 (2016)] and demonstrate excellent agreement with results from density functional theory calculations.   
\end{abstract}

\pacs{Valid PACS appear here}
\maketitle

The order of phonon modes, their polarization dependencies, and the corresponding structure of the reststrahlen bands are well understood for materials with orthorhombic and higher symmetries.\cite{VFSbook,Kittel2009,Wolfe89,GrundmannBook} The reststrahlen bands (Rubens and Nichols)\cite{RubensReststrahlen,Schaeferultrarot} permit accurate determination of long-wavelength polar lattice modes and their polarization dependencies.\cite{KruegerAP1928Reststrahlen,PalikJOSA1960reststrahlenlibrary} However, thus far structure and properties of the reststrahlen bands, and the order of phonon modes and their polarization dependencies in materials with monoclinic crystal symmetry remain unexplained. Material properties and the underlying physics for crystalline materials with monoclinic symmetry are gaining interest, for example, in electronic power devices, scintillators, high-power lasers, frequency stable laser local oscillators, light slowing and trapping devices, and optical quantum memory technologies.\cite{Higashiwakipssa2014,MikhailikJAP2005AWO4,KatoJPCS2005lasermat,CookPRL2015EuYSO,TurukhinPRL2001PrYSO,LvovskyNP2009OQM} In this letter, we explain the structure of the reststrahlen bands in monoclinic symmetry materials, and we identify the nature and physical origin of inner and outer phonon mode pairs. The results reported here may be useful for  understanding of physical properties associated with polar phonon mode coupling and propagation such as free charge carrier and thermal transport properties, and for correct identification of phonon modes by theory and experiment.

Born and Huang,\cite{Born54} in principle, laid out a formalism to derive the lattice dynamic properties in crystals with arbitrary symmetry. Solutions are categorized under different electric field $\mathbf{E}$ and dielectric displacement $\mathbf{D}$ conditions.\cite{VFSbook} $\mathbf{E}=0$ and $\mathbf{D}=0$  defines the transverse longitudinal (TO)  modes, $\omega_{\mathrm{TO},l}$, associated with dipole moment. $\mathbf{E} \ne 0$ but $\mathbf{D}=0$ defines the longitudinal optical (LO) modes, $\omega_{\mathrm{LO},l}$. $\mathbf{E}\ne 0$ and $\mathbf{D}\ne 0$ defines the so-called limiting frequencies $\omega(\mathbf{\alpha})_{l}$. While the application of the Born-Huang approach is straightforward to the calculation of the lattice dynamics in low symmetry materials, the underlying physics, hidden in high-symmetry materials due to degeneracies, remains to be discussed. For example, a generalization of the Lyddane-Sachs-Teller relation, a fundamental statement in solid state physics, was recently described for monoclinic and triclinic symmetry materials.\cite{SchubertPRL2016GLST} It was also observed that the so-called ``TO-LO rule'' for phonon modes in monoclinic $\beta-$Ga$_2$O$_3$ and Y$_2$SiO$_5$ is violated.\cite{SchubertPRB2016,MockPRB2018YSO} For orthorhombic and higher symmetry materials, all phonon mode displacements are polarized (directed) along three major linear Cartesian axes. For each axis, the sequence (order) of all phonon modes is such that a TO frequency is always followed exactly by one LO frequency with ascending wavelength (TO-LO rule). The thereby identified TO-LO pairs always possess the same displacement direction. For the previously investigated monoclinic materials $\beta-$Ga$_2$O$_3$ and Y$_2$SiO$_5$, for which complete sets of TO and LO modes determined from experiment are now available, the question remained unanswered as to which of the observed TO and LO modes form associated pairs, and whether such pair assignment can even be made for low-symmetry materials where all TO and LO modes differ in their directions within the monoclinic plane. The answer to this question is provided in this Letter.

The eigendielectric displacement vector summation approach describes the effect of polar vibrations onto the long-wavelength dependence of the dielectric function tensor, $\varepsilon$,  regardless of symmetry.\cite{SchubertPRB2016,SchubertPRL2016GLST,MockPRB2017CWO,MockPRB2018YSO} The approach is equivalent to the microscopic Born-Huang description of $N$ polar lattice vibrations in the harmonic approximation\cite{Born54,VFSbook} 

\begin{equation}\label{eq:EDVS}
\varepsilon=\varepsilon_{\infty}+\chi=\varepsilon_{\infty}+\sum^{N}_{l=1}\frac{A_{\mathrm{TO},l}(\mathbf{\hat{e}}_{\mathrm{TO},l}\otimes\mathbf{\hat{e}}_{\mathrm{TO},l})}{\omega^2_{\mathrm{TO},l}-\omega^2},
\end{equation}

\noindent where $A_{\mathrm{TO},l}$, $\omega_{\mathrm{TO},l}$, and $\mathbf{\hat{e}}_{\mathrm{TO},l}$ are amplitude, TO mode frequency, and unit eigen dielectric displacement vector of polar lattice mode $l$, respectively, $\varepsilon_{\infty}$ is the dielectric tensor contribution due to the combined vacuum permittivity and due to higher-frequency electronic dielectric polarization, $\otimes$ is the dyadic product, and $\omega$ is the time-harmonic frequency. The same statement can be formulated for $\varepsilon^{-1}$ with parameters for all LO modes exchanging all labels ``TO'' with ``LO" accordingly.\cite{MockPRB2018YSO}

Kuz'menko, Tishchenko and Orlov\cite{KuzmenkoJPCM1996} performed an eigenpolarization reflectance analysis for the monoclinic plane of $\alpha$-Bi$_2$O$_3$ introducing eigenpolarizations, $\mathbf{E_{\pm}}$, and corresponding wave propagation constants, $n_{\pm}$ (indices of refraction). Both $\mathbf{E_{\pm}}$ and $n_{\pm}$ derive from the time-harmonic photon-polariton dispersion relation, $\mathbf{k}(\hbar\omega)$, where the electromagnetic field wave vector is $\mathbf{k}=k_0(k_x,k_y,k_z)$

\begin{equation}\label{eq:Mawellwave}
\left[\varepsilon_{ij}-n^2\left(\delta-k_0^2k_ik_j\right)\right]E_j=0,
\end{equation}

\noindent where $\delta$ is the Kronecker symbol, $k_0=\frac{\omega}{c}$, and $c$ is the speed of light. For light at normal incidence to the $\mathbf{a-c}$ plane ($k_x=k_y=0$)

\begin{equation}
n_{\pm} = \sqrt{p \pm q},
\end{equation}

\noindent where

\begin{equation}
p = \varepsilon_{xx}+\varepsilon_{yy},\mbox{   }
q = \sqrt{(\varepsilon_{xx}-\varepsilon_{yy})^2+4\varepsilon_{xy}^2}.
\end{equation}

\noindent When $n_{\pm} \rightarrow  0$ it follows that $\det (\varepsilon_{ij}) \rightarrow  0$, and hence $\omega \rightarrow \omega_{\mathrm{LO},l}$. Consequently, 2 types of modes exist, one for $p-q=0$, and one for $p+q=0$. We refer to those as LO$_{-}$ and LO$_+$, respectively, and provide an explanation for their origins below. The normal incidence eigenpolarization ($\mathbf{E}_{\pm}$) reflectance Fresnel coefficients ($r_{\pm}$) can be expressed immediately, where $\mathbf{E}_{\pm}$ correspond to $n_{\pm}$, and are eigenvector solutions to Eq.~\ref{eq:Mawellwave} \cite{KuzmenkoJPCM1996}

\begin{equation}\label{eq:rpm}
r_{\pm}=\frac{n_{\pm}-1}{n_{\pm}+1}.
\end{equation}

Schubert, Tiwald and Herzinger\cite{SchubertPRB61_2000} identified conditions for bands of total reflection for high-symmetry orientations of surfaces cut from materials with orthorhombic and higher symmetry. Putting forward the same considerations as in Ref.~\onlinecite{SchubertPRB61_2000},\footnote{The imaginary parts in both sums $(r_{\pm}+r^{\star}_{\pm})$ and $(r_{\pm}-r^{\star}_{\pm})$ occurring in the numerator and denominator of $\sqrt{r_{\pm}r^{\star}_{\pm}}$, respectively, cancel regardless. Hence, when the real part of $n_{\pm}$ is zero, then the remaining ratio $\frac{n_{\pm}n^{\star}_{\pm}+1}{n_{\pm}n^{\star}_{\pm}+1} = 1$.} Eq.~\ref{eq:rpm} results in  

\begin{equation}\label{eq:totalRcondition}
\sqrt{r_{\pm}r^{\star}_{\pm}}=1 \Leftrightarrow Re\{n_{\pm}\} \leq 0.
\end{equation}  

\noindent where $^{\star}$ denotes the complex conjugate. The eigendielectric displacement vector summation approach can be used to identify the physics of the phonon modes within the monoclinic plane. We introduce $\mathbf{q}$ as a generalized displacement vector within the $\mathbf{a-c}$ plane:

\begin{equation}
\mathbf{q}=\sum^{N+2}_{l=1}\mathbf{q}_l,
\end{equation}

\noindent where we included the combined vacuum and higher electronic polarizability contributions, $\varepsilon_{xx,\infty}$ and $\varepsilon_{yy,\infty}$, as dyadic terms for $l=N+1$ and $l=N+2$, respectively.\footnote{Terms $\varepsilon_{xx,\infty}$, $\varepsilon_{yy,\infty}$, and $\varepsilon_{xy,\infty}$ can be represented by pole functions at an unspecified higher photon energy, $\omega_{\mathrm{TO,N+1}} = \omega_{\mathrm{TO,N+2}}$, far outside the long wavelength range and by unspecified amplitude parameters, $A_{\mathrm{N+1}}$, $A_{\mathrm{N+2}}$, and unit eigendisplacement vector orientations, $\alpha_{N+1}$, $\alpha_{N+1}$.} Displacement terms due to individual polar vibrations can be identified as follows

\begin{equation}
\mathbf{q}_l=\frac{A_{\mathrm{TO},l}}{\omega^2_{\mathrm{TO},l}-\omega^2}[\sin(2\alpha_{\mathrm{TO},l})\mathbf{\hat{e}}_x+\cos(2\alpha_{\mathrm{TO},l})\mathbf{\hat{e}}_y],
\end{equation}

\noindent where the factor 2 for the angular arguments accounts for the unidirectional polarizability of the phonon mode displacements, and $\hat{\mathbf{e}}_{\mathrm{TO},l}=\sin\alpha_{\mathrm{TO},l}\hat{\mathbf{x}}+\cos\alpha_{\mathrm{TO},l}\hat{\mathbf{y}}$. Then, it follows that magnitude $q=|\mathbf{q}|$ and the sum over all individual contributions, $p$,  
\begin{equation}
p=\sum^{N+2}_{l=1}|\mathbf{q}_l|=\sum^{N+2}_{l=1}\frac{A_{\mathrm{TO},l}}{\omega^2_{\mathrm{TO},l}-\omega^2},
\end{equation}
are identical with terms $p$ and $q$ in Eq.~\ref{eq:rpm}. Hence, $p$ and $q$ represent the dielectric polarizability within the monoclinic plane. Term $p$ is the total sum over all possible displacements produced by each individual lattice mode at a given wavelength including the quasi-static high-frequency contributions. Term $q$ takes the directional character of every mode within the monoclinic plane into account, and is a measure of the net magnitude of displacement at a given frequency.

\begin{figure*}[!tbp]
  \begin{center} 
  \includegraphics[width=.9\linewidth]{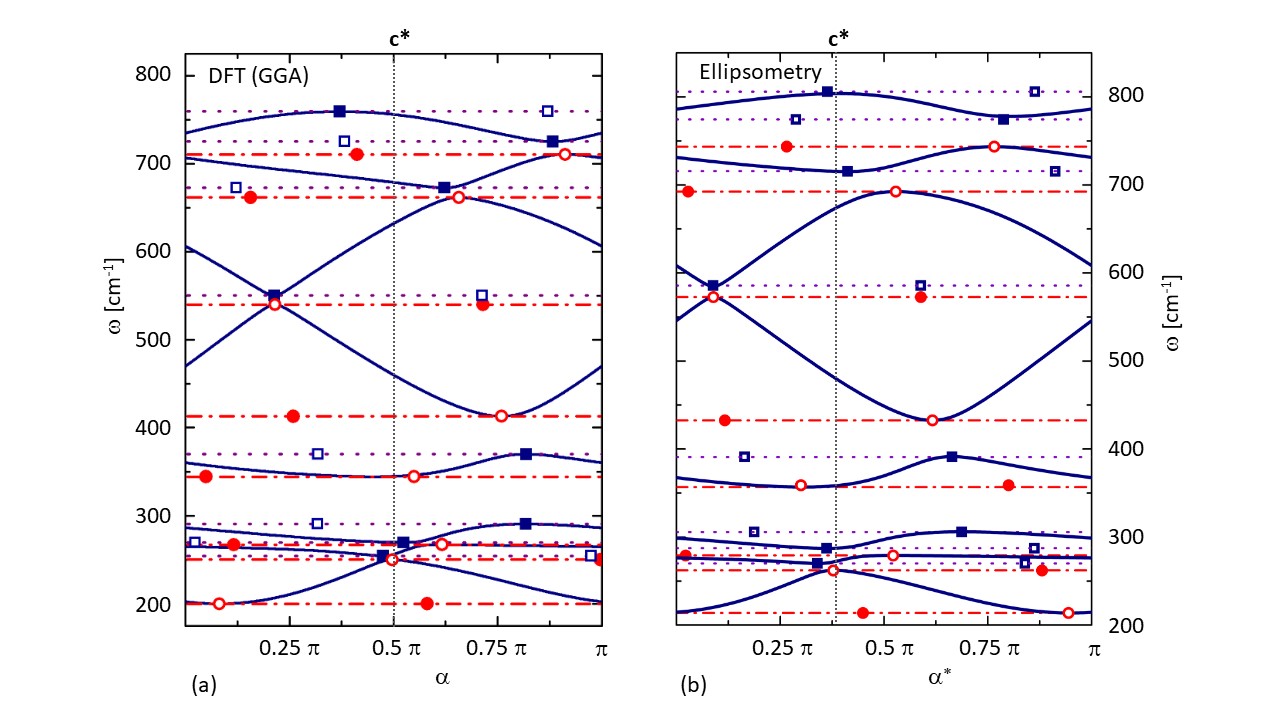}
    \caption{ Limiting long-wavelength mode frequencies $\omega(\mathbf{\alpha})_{l}$ (solid lines) of monoclinic symmetry $\beta$-Ga$_2$O$_3$ as a function of unit direction $\mathbf{\alpha}=\cos \alpha \mathbf{\hat{x}} + \sin\alpha\mathbf{\hat{y}}$ in the $\mathbf{a}$-$\mathbf{c}$ plane obtained from (a) density function theory calculations, and (b) generalized ellipsometry investigations. $\alpha=0: \mathbf{\alpha} \parallel \mathbf{a}$, $\alpha = \pi/2: \mathbf{\alpha} \parallel \mathbf{c}^{\star}$; $\mathbf{c}^{\star}$ is perpendicular to $\mathbf{a}$ within the $\mathbf{a}$-$\mathbf{c}$ plane, indicated by the vertical dashed line. The angular parameter $\alpha$ was offset by approximately -20$^{\circ}$ during experiment, $\alpha^{\star}=\alpha-20^{\circ}$. Indicated are frequencies and eigenvectors (closed symbols) and their normal directions (open symbols) of TO (red  circles), and LO modes (purple squares). Horizontal dash-dotted and dotted lines indicate frequencies of TO and LO modes, respectively. Parameters in (a) are calculated in this work, parameters in (b) are taken from experiment described in Ref.~\onlinecite{SchubertPRB2016}.}
    \label{fig:limitingfrequencies}
  \end{center}
\end{figure*}

If phonon mode broadening is ignored, which for the purpose of identifying the properties of phonon modes can be justified here, $p$ and $q$ are real-valued. It is important to recognize that $p$ can be both positive and negative, while $q$ is always positive. Two cases emerge from Eq.~\ref{eq:totalRcondition} 
\begin{equation}\label{eq:innermode}
\sqrt{r_{-}r^{\star}_{-}}=1 \Leftrightarrow p-q < 0,
\end{equation}
or
\begin{equation}\label{eq:outermode}
\sqrt{r_{+}r^{\star}_{+}}=1 \Leftrightarrow p+q < 0,
\end{equation}
In the first case, the net magnitude of displacement at a given frequency is larger than the sum over all possible displacements. This is the case for the onset of a phonon mode when no other resonance mode with negative displacement is occurring. At normal incidence, for ascending frequency, total reflection in $r_{-}$ occurs and forms a band, which stretches from a TO$_{-}$ to a LO$_{-}$. We refer to these modes as ``outer'' modes, which form pairs: [TO$_{j,-}$,LO$_{j',-}$], where $j$ and $j'$ index the occurrence of TO and LO modes, for example, with ascending frequency. In the second case, the net magnitude of displacement at a given frequency is smaller than the sum over all possible displacements. This is the case when a given phonon mode resonance produces negative displacement while a different polar resonance with negative displacement is already occurring. At normal incidence, bands of total reflection in $r_{+}$ stretch between pairs of frequencies of TO$_{+}$ and LO$_{+}$ modes, for ascending frequencies. We refer to these modes as ``inner'' modes, which form pairs: [TO$_{j',+}$,LO$_{j',+}$]. Pairs [TO$_+$,LO$_+$] fall within pairs of [TO$_-$,LO$_-$] because an inner mode occurs in a frequency region where another (outer) mode pair produces negative displacement. Hence, for example, with increasing frequency, phonon modes may appear with sequences $\dots$, TO$_{(-)}$, TO$_{(+)}$, $\dots$ and $\dots$, LO$_{(+)}$, LO$_{(-)}$, $\dots$, for example, as reported recently from experiment by generalized ellipsometry for Y$_2$SiO$_5$,\cite{MockPRB2018YSO} and $\beta$-Ga$_2$O$_3$.\cite{SchubertPRB2016} Multiple pairs [TO$_{j,+}$,LO$_{j',+}$] can fall within one pair [TO$_{-}$,LO$_{-}$], forming a structured phonon band. For example, the order and structure TO$_{8,-}$, TO$_{7,+}$, LO$_{8,+}$, TO$_{6,+}$, LO$_{7,+}$, LO$_{6,-}$ is observed within the respective 3 lowest-frequency $B_u$ symmetry TO and LO modes of $\beta$-Ga$_2$O$_3$.\cite{SchubertPRB2016}  Note that sequences with triple or more subsequent TO or LO modes, e.g., TO, TO, TO~$\dots$, or LO, LO, LO~$\dots$ cannot occur. 

The reststrahlen bands in the monoclinic plane are polarization dependent. For linearly polarized incident light, the reststrahlen bands are characterized by unpolarized bands of total reflection, polarized bands of total reflection, and intermediate bands with no total reflection. The latter occur outside bands of outer modes. Within inner mode pairs, [TO$_+$,LO$_+$], bands of unpolarized total reflection occur, since $r_+=r_-=1$. Hence, at normal incidence the $\mathbf{a-c}$ plane of a monoclinic crystal is totally reflective within all inner bands, regardless of light polarization. Within spectral regions inside outer mode pairs not overlaid by inner mode pairs, [TO$_-$,LO$_-$]$\cap$[TO$_+$,LO$_+$], total reflection occurs only for one linear polarization state. The angle depends on the wavelength, hence, we refer to these bands as polarization (or angular) dependent bands. Narrow lines of total reflection form, which  begin and end at frequencies and directions of the modes in [TO$_-$,LO$_-$]$\cap$[TO$_+$,LO$_+$]. The angles at which the bands begin and end can be directly read from the directions of the eigenvectors,\footnote{Note that in the case of no broadening, the eigenvectors do not have imaginary components and are thus always linearly polarized, regardless of wavelength.} $\varphi_{\pm}=\tan^{-1}(E_{\pm,y}/E_{\pm,x})$ at begin and end of the polarization dependent bands. It can be shown analytically that when $\omega \rightarrow \omega_{s_{j,-}}$, with $s \in \{ \mathrm{TO}, \mathrm{LO}\}$

\begin{equation}
\tan^{-1} \left( \varphi_{-} \right)=\alpha_{\mathrm{s}_{j,-}}.
\end{equation}

\noindent Likewise, when $\omega \rightarrow \omega_{s_{j,+}}$,

\begin{equation}
\tan^{-1} \left( \varphi_{+} \right)=\alpha_{\mathrm{s}_{j,+}} + \pi/2.
\end{equation}

\noindent Hence, bands begin and end at outer mode frequencies, and mark the direction of the phonon modes. Bands connecting to inner mode frequencies identify directions perpendicular to the phonon mode directions. 

The limiting frequencies, $\omega(\alpha)_{l}$, depend on the direction of a unit vector within the $\mathbf{a-c}$ plane, $\mathbf{\alpha}=\cos{\alpha}\mathbf{\hat{x}}+\sin{\alpha}\mathbf{\hat{y}}$. In crystals with orthorhombic and higher symmetry, frequencies $\omega(\alpha)_{l}$ are bound within associated TO-LO pairs: $\omega_{\mathrm{TO},l} \leq \omega(\alpha)_{l} \leq \omega_{\mathrm{LO},l}$. Hence, TO and LO frequencies, and their band association $l$, for example, in density functional theory (DFT) calculation results, can be identified from  minima and maxima, and band index $l$ of calculated $\omega(\alpha)_{l}$, respectively. However, this assignment can be incorrect for crystals with monoclinic symmetry as a direct consequence of the phonon order and structure discussed above. Because no solution for $\omega(\alpha)_{l}$ exists within bands of total reflection, the allowed frequency regions are confined to frequency regions [TO$_-$,LO$_-$]$\cap$[TO$_+$,LO$_+$]. Hence, minimum - maximum bounds for $\omega(\alpha)_{l}$ can be ($i$) $\omega_{\mathrm{TO}_{j,-}}$ - $\omega_{\mathrm{TO}_{j',+}}$, ($ii$) $\omega_{\mathrm{LO}_{j,+}}$ - $\omega_{\mathrm{TO}_{j',+}}$, ($iii$) $\omega_{\mathrm{LO}_{j,+}}$ - $\omega_{\mathrm{LO}_{j',-}}$, and ($iv$) $\omega_{\mathrm{TO}_{j,-}}$ - $\omega_{\mathrm{LO}_{j',-}}$.

Figure~\ref{fig:limitingfrequencies} depicts frequencies $\omega(\mathbf{\alpha})_{l}$ for the $\beta$-Ga$_2$O$_3$ $\mathbf{a}$-$\mathbf{c}$ plane as a function of $\alpha$ obtained by DFT (a) and from ellipsometry (b). DFT results were computed at the $\Gamma$ point of the Brillouin zone as described in Ref.~\onlinecite{SchubertPRB2016} for the fully relaxed unit cell of $\beta-$Ga$_2$O$_3$, and using the same method including the general gradient approximation (GGA) density functional of Perdew-Burke-Ernzerhof\cite{PBE1996} as in Ref.~\onlinecite{Mock_2017Ga2O3}. TO mode frequencies were taken directly from the $\Gamma$-point calculations. Frequencies $\omega(\mathbf{\alpha})_{l}$ with $B_u$ symmetry were obtained by setting a small displacement from the $\Gamma$-point, so that the entire $\mathbf{a-c}$ plane was probed with a step of 0.1$^\circ$. The extrema of the dispersion curves were marked as LO modes if they did not coincide with previously identified TO modes. Mode frequencies $\omega(\mathbf{\alpha})_{l}$ in Fig.~\ref{fig:limitingfrequencies}(b) are obtained from the observed zero crossings in the real part of $\varepsilon_{\alpha\alpha}$, determined from measurements. An excellent agreement between theory and experiment is noted. Eight bands of $\omega(\mathbf{\alpha})_{l}$ occur, as expected from the observed 8 TO and 8 LO modes. However, the maxima of frequencies $\omega(\mathbf{\alpha})_{l}$ are not always identical with LO modes. Likewise, the minima are not always identical with TO modes. Specifically, when an outer mode TO-LO pair is not disrupted by inner mode pairs, minima and maxima for $\omega(\mathbf{\alpha})_{l}$ are identical with TO and LO modes. However, when inner mode pairs exist, $\omega(\mathbf{\alpha})_{l}$ reveal gaps with anti-crossing behavior. A similar behavior was noted from results of DFT calculations for phonon modes in monoclinic-symmetry CsSnCl$_3$.\cite{HuangPRB2015CsSnCl3DFT} When inner mode pairs exist, $\omega(\mathbf{\alpha})_{l}$ can be limited by two TO modes, or two LO modes, or the upper limit can be a TO mode while the lower limit can be a LO mode. This result is important for correct identification and proper assignment of TO and LO mode character and their correct eigenvectors from DFT calculations.

\begin{figure}[!tbp]
  \begin{center} 
  \includegraphics[width=0.9\linewidth]{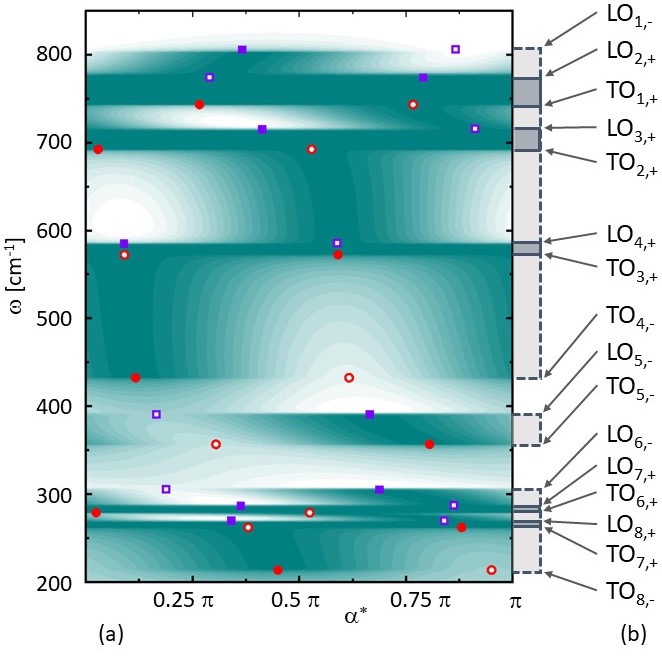}
    \caption{ (a) Reflectance color-density plot (10 equally spaced density scales between dark olive for $R_{\alpha^{\star}}=1$ and white for $R_{\alpha}=0$) rendering the linearly polarized reststrahlen bands for the $\beta$-Ga$_2$O$_3$ $\mathbf{a}$-$\mathbf{c}$ plane as a function of $\alpha^{\star}$ and symbols as defined in Fig.~\ref{fig:limitingfrequencies}. The incident polarization is parallel to $\mathbf{\alpha^{\star}}$. (b) 3 pairs of outer modes, and 5 pairs of inner modes occur in $\beta$-Ga$_2$O$_3$, forming 3 phonon bands with phonon mode order matching the previously observed order of $B_u$-symmetry TO and LO frequencies in Ref.~\onlinecite{SchubertPRB2016}.}
    \label{fig:eigendisplacement}
  \end{center}
\end{figure}

Figure~\ref{fig:eigendisplacement} depicts the linearly polarized reststrahlen ($R_{\alpha^{\star}}$) bands of $\beta$-Ga$_2$O$_3$ in the monoclinic plane (Fig.~\ref{fig:eigendisplacement}(a)), and a schematic of the associated phonon bands (Fig.~\ref{fig:eigendisplacement}(b)). $R_{\alpha^{\star}}$ is calculated using the eigendielectric displacement vector summation approach as described in Ref.~\onlinecite{SchubertPRB2016}. The influence of broadening is ignored here for clarity, and all broadening parameters are set to zero. The diagrams show frequencies and directions of all TO and LO modes (solid symbols).  TO and LO mode frequencies are indicated by horizontal lines. Indicated are further the directions normal to all TO and LO modes within the monoclinic plane (open symbols). A total of 8 TO and 8 LO modes is observed from experiment. The modes form 5 spectral regions of polarization-independent total reflection and 8 regions of polarization-dependent total reflection. The reststrahlen bands separate into 3 structured frequency regions within which a frequency can be found and at least one linear polarization can be found where $R_{\alpha}=1$. These regions are bound by a thereby associated outer mode TO-LO pair. The frequency regions of polarization-independent total reflection are bound by associated inner mode TO-LO pairs. Note that in regions of polarization-dependent total reflection, total reflectance begins and ends at directions which are normal to the eigendirections of the TO and/or LO modes involved in the band. When the polarization-dependent total reflection begins and/or ends at an outer mode, it begins and/or ends at the outer mode eigenvector direction. The resulting order of outer and inner phonon mode pairs is depicted in Fig.~\ref{fig:eigendisplacement}(b). Note that a structured band (an outer mode pair with embedded inner mode pairs) always begins and ends with sequences of double LO and double TO occurrences, respectively. 

An eigendielectric displacement vector summation approach was used to reveal the existence of inner and outer polar phonon mode pairs in materials with monoclinic symmetry. We thereby explained the unusual frequency order and the polarization-dependent structure of the reststrahlen bands, and demonstrated our findings for the case of monoclinic $\beta$-Ga$_2$O$_3$. The directional limiting modes within the Born-Huang approach are bound to within outer mode regions not occupied by inner mode pairs. Hence, an unusual phonon mode order can occur where both lower-frequency as well as upper-frequency limits for the directional modes can be both transverse and/or longitudinal modes. An excellent agreement was found for all statements made for $\beta$-Ga$_2$O$_3$ with results from density functional theory calculations.

This work was supported in part by the National Science Foundation under award DMR 1808715, by Air Force Office of Scientific Research under award FA9550-18-1-0360, and by the Nebraska Materials Research Science and Engineering Center under award DMR 1420645. We acknowledge support from the  Swedish Energy Agency under award P45396-1, the Swedish Research Council VR under award No. 2016-00889, the Swedish Foundation for Strategic Research under Grant Nos. FL12-0181, RIF14-055, and EM16-0024, and the Swedish Government Strategic Research Area in Materials Science on Functional Materials at Link{\"o}ping University, Faculty Grant SFO Mat LiU No. 2009-00971. M.~S. acknowledges the University of Nebraska Foundation and the J.~A.~Woollam~Foundation for financial support. 

\bibliography{CompleteLibrary}
\end{document}